\documentclass[aps,prl,twocolumn,superscriptaddress,showpacs,letter]{revtex4-1}

\usepackage{graphicx}
\usepackage{dcolumn}
\usepackage{bm}
\usepackage{cancel}
\usepackage{color}

\newcommand{\ac}{A$_3$C$_{60}$}

\newcommand{\AFS}{A$_{x}$Fe$_{2-y}$Se$_2$}
\newcommand{\KFS}{K$_{x}$Fe$_{2-y}$Se$_2$}
\newcommand{\RFS}{Rb$_{0.8}$Fe$_{2}$(Se$_{1-z}$S$_z$)$_2$}
\newcommand{\RFSa}{Rb$_{0.8}$Fe$_{2}$Se$_2$}
\newcommand{\RFSb}{Rb$_{0.8}$Fe$_{2}$SeS}
\newcommand{\RFSc}{Rb$_{0.8}$Fe$_{2}$S$_2$}
\newcommand{\RFSins}{Rb$_{0.8}$Fe$_{1.5}$S$_2$}
\newcommand{\KFSins}{K$_{0.85}$Fe$_{1.54}$S$_2$}
\newcommand{\ef}{$E_F$}

\newcommand{\kz}{$k_z$}
\newcommand{\kx}{$k_x$}
\newcommand{\tc}{$T_C$}

\newcommand{\dxy}{$d_{xy}$}
\newcommand{\dxz}{$d_{xz}$}
\newcommand{\dyz}{$d_{yz}$}

\newcommand{\mub}{$\mu_{\tiny{\textrm{B}}}$}
\newcommand{\g}{$\Gamma$}

\begin{document}

\title{Bandwidth and Electron Correlation-Tuned Superconductivity in \RFS}

\author{M. Yi}
\email{mingyi@berkeley.edu}
\affiliation{Department of Physics, University of California Berkeley, Berkeley, CA 94720, USA}
\author{Meng Wang}
\email{wangm@berkeley.edu}
\affiliation{Department of Physics, University of California Berkeley, Berkeley, CA 94720, USA}
\author{A. F. Kemper}
\affiliation{Computational Research Division, Lawrence Berkeley National Lab, Berkeley, CA 94720, USA}
\author{S.-K. Mo}
\author{Z. Hussain}
\affiliation{Advanced Light Source, Lawrence Berkeley National Lab, Berkeley, CA 94720, USA}
\author{E. Bourret-Courchesne}
\affiliation{Materials Science Division, Lawrence Berkeley National Laboratory, Berkeley, CA 94720, USA}
\author{A. Lanzara}
\affiliation{Department of Physics, University of California Berkeley, Berkeley, CA 94720, USA}
\affiliation{Materials Science Division, Lawrence Berkeley National Laboratory, Berkeley, CA 94720, USA}
\author{M. Hashimoto}
\affiliation{Stanford Synchrotron Radiation Lightsource, SLAC National Accelerator Laboratory, Menlo Park, CA 94025, USA}
\author{D. H. Lu}
\affiliation{Stanford Synchrotron Radiation Lightsource, SLAC National Accelerator Laboratory, Menlo Park, CA 94025, USA}
\author{Z.-X. Shen}
\affiliation{Stanford Institute of Materials and Energy Sciences, Stanford University, Stanford, CA 94305, USA}
\affiliation{Departments of Physics and Applied Physics, and Geballe Laboratory for Advanced Materials, Stanford University, Stanford, CA 94305, USA}
\author{R. J. Birgeneau}
\affiliation{Department of Physics, University of California Berkeley, Berkeley, CA 94720, USA}
\affiliation{Materials Science Division, Lawrence Berkeley National Laboratory, Berkeley, CA 94720, USA}
\affiliation{Department of Materials Science and Engineering, University of California, Berkeley, CA 94720, USA}

\date{\today}

\begin{abstract}
We present a systematic angle-resolved photoemission spectroscopy study of the substitution-dependence of the electronic structure of \RFS~(z = 0, 0.5, 1), where superconductivity is continuously suppressed into a metallic phase. Going from the non-superconducting \RFSc~to superconducting \RFSa, we observe little change of the Fermi surface topology, but a reduction of the overall bandwidth by a factor of 2. Hence for these heavily electron-doped iron chalcogenides, we have identified electron correlation as explicitly manifested in the quasiparticle bandwidth to be the important tuning parameter for superconductivity, and that moderate correlation is essential to achieving high \tc.
\end{abstract}


\maketitle


\begin{figure}
\includegraphics[width=0.5\textwidth]{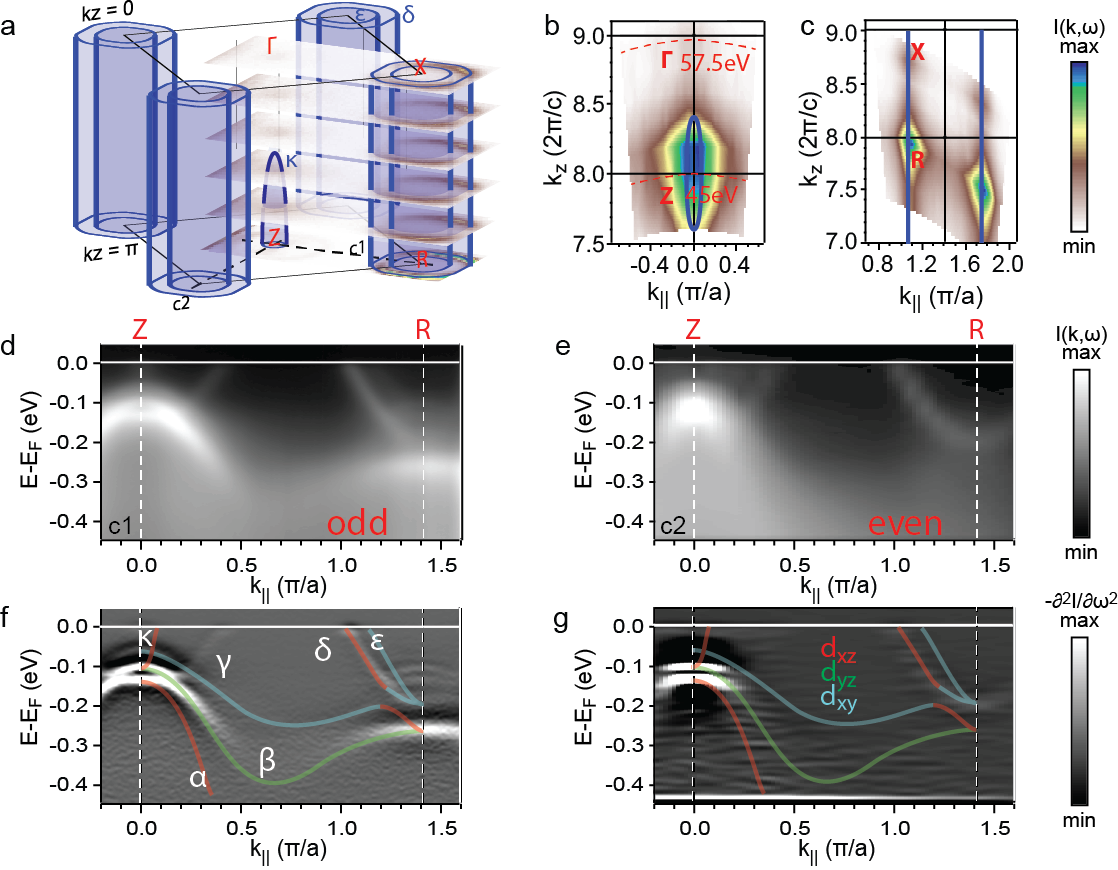}
\caption{\label{fig:fig1}Electronic structure of RFS. (a) Three-dimensional illustration of the Fermi surface (FS) with measured FS slices shown from \kz~= 0 to \kz~= $\pi$, taken with 45 eV to 57.5 eV photons in steps of 2.5 eV. Brillouin zone (BZ) notations pertaining to the 2-Fe unit cell are used. \kz~map of (b) the electron pocket near the Z point and (c) the electron pocket along the BZ corner, taken along the \g-X direction with even polarization. (d)-(e) Spectral images along cuts c1 and c2 as labeled in (a), with odd/even polarizations with respect to the cut direction, respectively. (f)-(g) Second energy derivatives for (d)-(e), respectively. Dominant orbital characters are marked with color.}
\end{figure}

In the current study of iron-based superconductors, one of the challenges is to understand the superconductivity (\tc~$\sim$~30 K) in the heavily electron-doped iron chalcogenides \AFS~(A = alkali metal) despite their lack of the ubiquitous Fermi surface (FS) nesting conditions previously thought to be a key for the iron-pnictide superconductivity~\cite{Zhang2011,Mou2011,Qian2011}. Moreover, an antiferromagnetically ordered insulating phase with a spin $S$ = 2 and moment as large as 3.3 \mub~\cite{Bao2011} has been discovered to exist in proximity to superconductivity in \AFS. Subsequently, many theories have been proposed to understand the superconductivity in \AFS~from a strong coupling approach~\cite{Yu2011,Zhou2011,Craco2011}, where superconductivity appears in proximity to a Mott phase. Thereby, it becomes important to determine experimentally the key tuning parameters for superconductivity in this family. 

However, one of the initial challenges has been to control the stoichiometry of the material composition in the growth process, preventing a systematic way to tune the \tc. More recently, it has been shown that substitution of selenium by sulfur can tune and suppress \tc~continuously~\cite{Lei2011,Li2011c}, offering a pathway for systematically studying the emergence of superconductivity in this family. 

In this Letter, we use angle-resolved photoemission spectroscopy (ARPES) to study the \RFS~(z = 0, 0.5, 1) series, where the replacement of selenium by sulfur progressively turns a 32 K superconductor into a non-superconducting metallic phase. In the electronic structure, we observe minimal changes in the FS topology, accompanied by a small change of the total charge carrier concentration. More significantly, from \RFSc~to \RFSa, the overall quasiparticle bandwidth is observed to decrease by a factor of 2, signaling a dramatic change in electron correlations. This can be understood by considering the bigger size of the selenium atoms compared to sulfur, which expands the lattice, hence increasing the overall electron correlations. Our results show that for the alkali-metal doped iron chalcogenides, electron correlations as controlled by the bandwidth, rather than charge carrier doping or FS topology, is the important tuning parameter for superconductivity, and moderate correlation is essential to achieving high \tc.

High quality single crystals with nominal compositions of \RFSa~(RFSe), \RFSb~(RFSeS), and \RFSc~(RFS), were grown using the flux method~\cite{Wang2014,Wang2015}. Their actual compositions determined by energy-dispersive x-ray spectroscopy are Rb$_{0.67}$Fe$_{1.65}$Se$_2$, Rb$_{0.77}$Fe$_{1.64}$Se$_{1.03}$S$_{0.97}$, and Rb$_{0.75}$Fe$_{1.85}$S$_2$, respectively. ARPES measurements were performed at beamline 10.0.1 of the Advanced Light Source and beamline 5-4 of the Stanford Synchrotron Radiation Lightsource, using R4000 electron analyzers, with an energy resolution better than 15 meV and angular resolution of 0.3$^{\circ}$. All samples were cleaved \emph{in situ} and measured at 30 K with 45 eV photos unless noted otherwise, in an ultra-high vacuum with a base pressure better than 4$\times$10$^{-11}$ torr. Density functional theory calculations were performed with the PBE exchange functional~\cite{Perdew1996} using the full potential (linearized) augmented plane-wave method as implemented in the WIEN2k package~\cite{Blaha2001}, and using a k-mesh of $10\times10\times10$ in the primitive unit cell basis. The calculations were based on experimentally determined structure parameters~\cite{Wang2015,Pomjakushin2012}.

\begin{figure}
\includegraphics[width=0.5\textwidth]{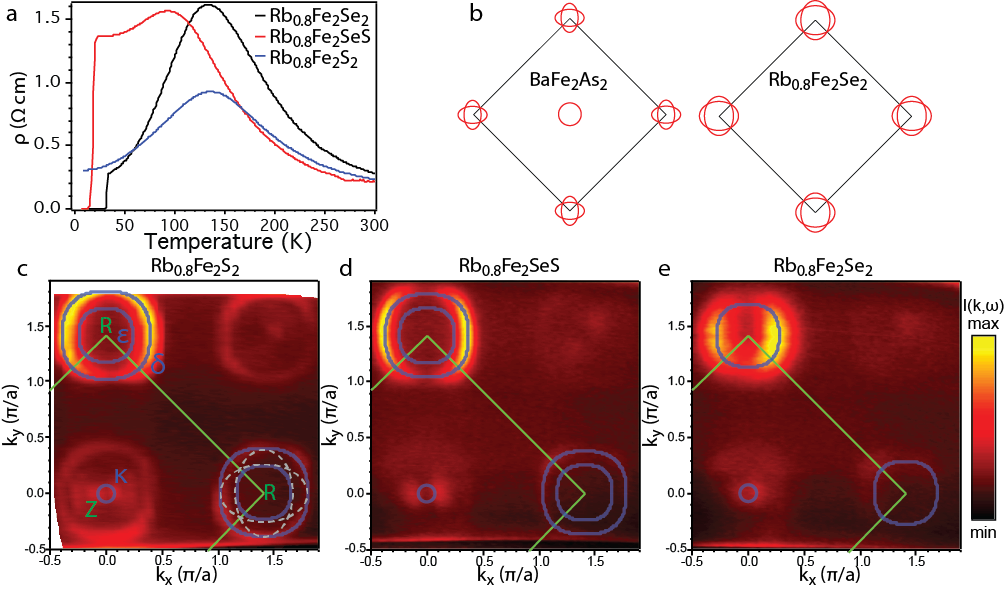}
\caption{\label{fig:fig2}Substitution-dependence of FSs. (a) Resistivity data for RFSe (black), RFSeS (red), and RFS (blue). (b) Schematic contrasting the RFSe FS topology with that of BaFe$_2$As$_2$. FSs (\kz~near $\pi$) of (c) RFS, (d) RFSeS, and (e) RFSe. Green lines mark the boundaries of the 2-Fe BZ. Blue lines are the FS outlines. Gray dotted lines in (b) illustrate the original intersecting elliptical pockets before hybridization. Light polarization is along the \kx~direction. }
\end{figure}

The measured electronic structure of RFS is summarized in Fig.~\ref{fig:fig1}. The most conspicuous feature of the FS is a large pocket ($\delta$) at the Brillouin zone (BZ) corner, which is two-dimensional as revealed by the \kz-dependent maps (Fig.~\ref{fig:fig1}a,c). In addition, there is a small and highly three-dimensional closed pocket ($\kappa$) centered around the Z point, as can be seen in its dispersive nature in the \kz~map (Fig.~\ref{fig:fig1}b). From the band structure measured along the high symmetry direction Z-R (Fig.~\ref{fig:fig1}d-g), the small pocket at Z and the large cylindrical pockets at the BZ corner are all revealed to be electron-like. In addition, we observe three hole-like bands below the Fermi level (\ef) at the BZ center. Hence we have identified the five familiar Fe $3d$ bands (Fig.~\ref{fig:fig1}f-g). However, closer examination of both the high symmetry bands (Fig.~\ref{fig:fig1}f) and FSs (Fig.~\ref{fig:fig2}b) reveal extra features with much weaker intensity, which are exact copies of the bands after a $\bf{q}$ = ($\pi$,$\pi$) folding between the BZ center and corner. This $\bf{q}$ matches an ordering of the K atoms on the cleaved surface in the analogous compound, \KFS~\cite{Wang2012b}, and is likely of a similar origin here. Furthermore, using polarization studies~\cite{SI}, we can identify the orbital characters of the hole bands to be dominantly \dxz, \dyz, and \dxy~for $\alpha$, $\beta$, and $\gamma$, respectively, and for the electron bands \dxz~and \dxy~for $\delta$ and $\epsilon$, respectively, consistent with that found for other iron chalcogenides~\cite{Yi2013,Yi2015}.

The electronic structure of RFS~is qualitatively similar to the reported electronic structure of \AFS~\cite{Zhang2011,Mou2011,Qian2011,Yi2013}, and consists of only large electron pockets in contrast to other iron-based high temperature superconductors which have compensated electron and hole Fermi pockets, as shown for BaFe$_2$As$_2$ (Fig.~\ref{fig:fig2}b). Next, we study the detailed evolution from RFS to RFSe. Resistivity measurements show that RFS does not superconduct and remains a metal at low temperatures, while RFSeS and RFSe have \tc s of 20 K and 32 K, respectively (Fig.~\ref{fig:fig2}(a))~\cite{Wang2015}. From RFS to RFSe, little change occurs in the FS topology (Fig.~\ref{fig:fig2}c-e), which is dominated by large electron pockets at the BZ corner. One difference is that, while the two electron pockets in RFS are well separated, they become closer in size in RFSeS and merge into two nearly degenerate pockets in RFSe. The complete separation of two electron pockets comes from hybridization of two crossing electron ellipses (Fig.~\ref{fig:fig2}b). This hybridization is rarely seen in other iron-based compounds, and can only occur in the presence of glide mirror symmetry breaking (such as difference in the chalcogen height above and below the iron plane) or spin-orbit coupling, the latter having been reported in NaFeAs~\cite{Borisenko2014}. However, here the splitting is much larger than that observed in NaFeAs. Its source remains as an open question. From the FS volume enclosed, we can determine the total charge carrier concentration from RFS to RFSe to be 0.36 e$^-$/2Fe, 0.34 e$^-$/2Fe, and 0.27 e$^-$/2Fe, indicating a small variation in electron doping. As sulfur substitution for selenium is an isovalent process, we do not expect the charge carrier level to change significantly. The small change here may indicate a variation of the Rb content or Fe valency.


\begin{figure}
\includegraphics[width=0.48\textwidth]{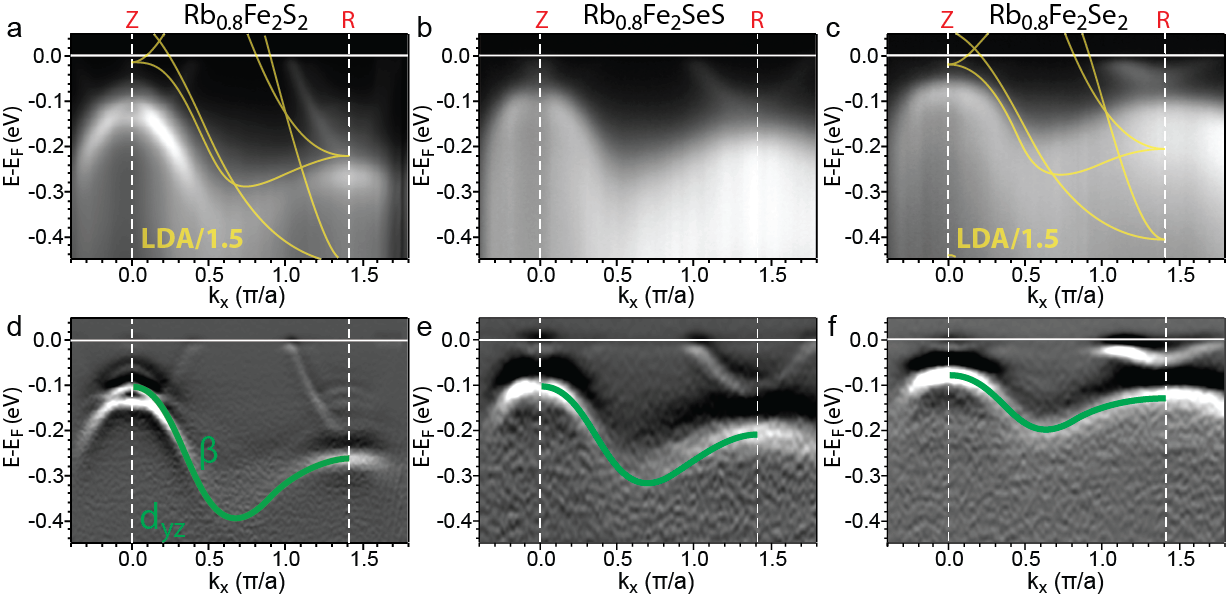}
\caption{\label{fig:fig3}Substitution-dependence of bandwith. (a)-(c) Measured spectral images along the high symmetry direction Z-R on RFS, RFSeS, and RFSe, respectively. Yellow lines are calculated LDA bands for RbFe$_2$S$_2$ and RbFe$_2$Se$_2$ along Z-R renormalized by a factor of 1.5 for comparison. (d)-(e) Second energy derivatives of (a)-(c). The green lines outline the \dyz-dominated band.}
\end{figure}

\begin{figure}
\includegraphics[width=0.5\textwidth]{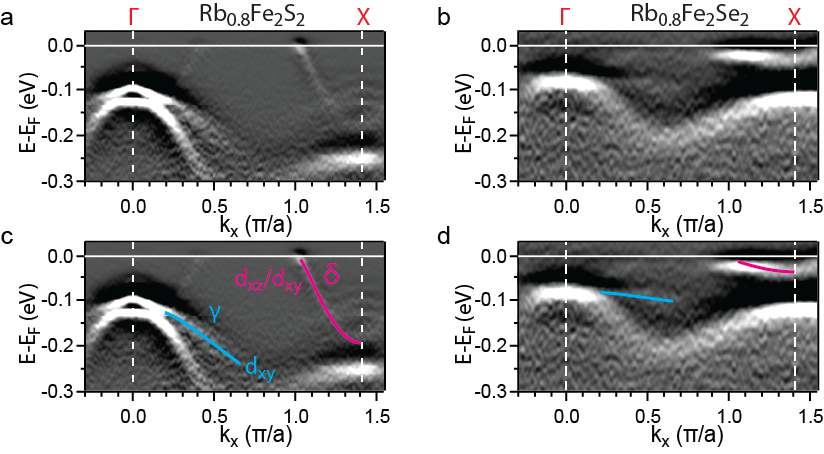}
\caption{\label{fig:fig4}Substitution-dependence of the \dxy~band. (a)-(b) Second energy derivatives of spectral images taken with 35 eV photons along the high symmetry direction \g-X for RFS~and RFSe. (c)-(d) The same images from (a)-(b) with the \dxy~band marked in blue, and the electron band marked in magenta, which has mixed \dxz~and \dxy~character.}
\end{figure}

Next, we compare the band structure across the substitution series from RFS to RFSe by comparing the high symmetry direction Z-R in the same energy window (Fig.~\ref{fig:fig3}). Remarkably, we see that while their structures are qualitatively similar, the bandwidths of all the observable bands are drastically decreased. As an example, we highlight in green the \dyz~band, which can be observed in its entirety. Its bandwidth, $W_{yz}$, decreases from 270 meV in RFS to 130 meV in RFSe, by roughly a factor of 2. In addition to this overall bandwidth reduction, we see that the \dxy~hole band (marked in blue in Fig.~\ref{fig:fig4}c-d) becomes selectively flatter compared to the \dyz~band, with its band velocity, $v_{xy}$, decreasing by a factor of 6.4, indicating that the \dxy~orbital is increasingly more localized compared to the other orbitals. Similarly at the X point, the $\delta$~electron band also flattens, as indicated by a 3-fold increase of the effective mass extracted from a parabolic fit to the dispersion (Fig.~\ref{fig:fig4}c-d). Here we note that this band is a mixture of both \dxz~and \dxy~orbitals (Fig.~\ref{fig:fig1}f-g)~\cite{Yi2015}, hence possibly shows a behavior that is intermediate of \dxz/\dyz~and \dxy.


Before we discuss the systematic changes with substitution, we would like to note that these \RFS~compounds do have similar phase separation as found for the \AFS~materials, where there always exists a block antiferromagnetic 245 phase that is phase separated from the metallic phase~\cite{Chen2011,Pomjakushin2012}, as confirmed by neutron scattering measurements~\cite{Wang2015}. However, the phases that we have probed near \ef~are clearly metallic in all three compounds. This is made possible due to the insulating nature of the 245 phase, which does not contribute photoemission intensity near \ef~\cite{Chen2011,Yi2013}, giving us the opportunity to probe directly the metallic phase that is continuously connected to the superconducting phase in \AFS. In addition, we would like to distinguish these metallic phases from the magnetically ordered insulating 234 phase reported in \KFSins~and \RFSins~\cite{Zhao2012,Wang2015a}, where there exists ordered iron vacancy. Here, the metallicity is due to the almost full iron content~\cite{Texier2012}.

\begin{figure}
\includegraphics[width=0.45\textwidth]{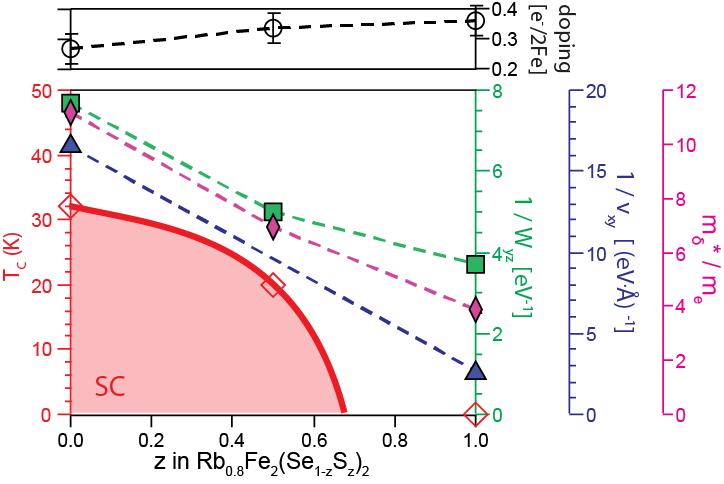}
\caption{\label{fig:fig5}Various parameters plotted as a function of S content, $z$: carrier doping estimated from FS volume (black circles), inverse of the \dyz~bandwidth, $W_{yz}$ (green squares), inverse of the \dxy~band slope, $v_{xy}$, between \kx~= 0.3 to 0.7 ($\pi$/a) (blue triangles), the effective mass of the $\delta$ electron band divided by the free electron mass, $m^*_\delta/m_e$ (magenta diamonds), and \tc~(open red diamonds).}
\end{figure}

Since \tc~changes from non-existent in RFS~to 32 K in RFSe, it is important to identify the parameter dictating superconductivity in this system. From RFS~to RFSe, clearly little change occurs in the FS topology while the bandwidth is remarkably reduced, indicating a change in the electron-electron correlations. To gauge the strength of electron correlations, we have performed LDA calculations for RbFe$_2$S$_2$ and RbFe$_{2}$Se$_2$. The calculated bands (renormalized by a factor of 1.5) are overlaid in Fig.~\ref{fig:fig3}a,c, where interestingly the one-electron bandwidth for \dyz~shrinks from 0.42 eV to 0.37 eV, but by a mere factor of 1.1, compared to the factor of 2 for measured bandwidth. In other words, the renormalization factor increases from 1.6 to 2.8, indicating a strong enhancement of electron correlation with growing Se content. It is also interesting to point out that, a simple overall renormalization of the LDA bands is not sufficient to obtain satisfactory agreement with the data  (Fig.~\ref{fig:fig3}). An additional momentum-dependent shift up (down) of the LDA electron (hole) bands is needed. This behavior has been reported for iron-pnictides~\cite{Yi2009,Brouet2013}, and has been ascribed to the effect of interband interactions~\cite{Ortenzi2009,Zhai2009}. Here, this effect is observed to increase with overall electron correlations, where in the RFSe~case, the bottom of the \dxy~electron band is raised completely above the \dxy~hole band top. 

It has been shown that both carrier doping and lattice parameters modulate electron correlation in the iron-based superconductors~\cite{Nakajima2014,Ye2014}, paralleling the effects of filling-controlled and bandwidth-controlled Mott transition. Here, the small decrease of electron doping with increasing Se content is consistent with the increase in electron correlation as the loss of electron doping brings the material towards half filling. However, the change of 0.045 e$^-$/Fe from RFS~to RFSe~is much too small to account for a bandwidth change of a factor of 2, when compared with the other iron-based materials~\cite{Ye2014}, where Co-doping in LiFeAs of 0.1 e$^-$/Fe, for example, reduces the bandwidth only by a factor of 1.3. Next, we consider the lattice parameter changes. Sulfur atoms are much smaller than selenium. Hence sulfur substitution naturally shrinks the lattice parameters. The lattice constant $a$ of the metallic phase indeed decreases from 3.832 \AA~in RFSe~\cite{Pomjakushin2012}~to 3.692 \AA~in RFS~\cite{Wang2015}, consistent with the more localized behavior in the expanded lattice of RFSe~compared to the more itinerant behavior in RFS, as lattice expansion reduces electron hopping~\cite{Zhu2010}. Moreover, the bigger selenium atom increases the anion height, hence rendering the mostly in-plane \dxy~orbital more localized than the other orbitals as the hopping between \dxy~orbitals involves the $p_x$/$p_y$ orbitals of the chalcogen, resulting in the observed increasingly selective band renormalization of the \dxy~orbital (Fig.~\ref{fig:fig4}). This naturally connects to the observation of the orbital-selective Mott crossover in superconducting \AFS~\cite{Yi2013}, and highlights the importance of moderate electron correlations and bandwidth renormalization for achieving high \tc. To quantitatively illustrate this, we plot in Fig.~\ref{fig:fig5} $1/W_{yz}$, $1/v_{xy}$, and $m^*_\delta/m_e$, which all scale with $1/Z$, where $Z$ is the quasiparticle spectral weight, which approaches zero towards a Mott localization. Here we see that towards RFSe, where the highest \tc~occurs, $1/Z$ as estimated from all the bands increases, indicative of increasing electron correlations. It is interesting to point out a striking similarity between this system and the \ac~alkali-doped fullerides, which is a narrow-band system that exhibits superconductivity in proximity to a Mott insulating state~\cite{Gunnarsson2004,Chen1991}. Interestingly, the bandwidth of \ac~compounds also narrows with the expansion of the lattice. While the \tc~initially rises with this narrowing of the bandwidth, it eventually reaches a maximum before lowering again, forming a dome-like structure while the bandwidth continues to shrink together with the increase of the density of states at \ef, towards the metal-insulator transition (MIT)~\cite{Ganin2010}. For the \AFS~system, further isovalent substitution of Se by the bigger Te atoms has indeed been revealed to reduce \tc, demonstrating a superconducting dome similar to that of the \ac~\cite{Gu2012}, showing that moderate electron correlation as seen in the narrowing of the quasiparticle bandwidth is important for superconductivity until a threshold is reached where the system approaches the MIT~\cite{Capone2002}.

Finally, we would like to discuss the effect of electron correlation on superconductivity in \RFS~in the context of iron-based superconductors. It has been nicely shown that electron correlation is reflected in an overall bandwidth change across several iron-based superconductor phase diagrams~\cite{Ye2014}. However, in the 111 and 122 systems (even including the isovalent P-substitution), the relative sizes of the electron and hole FSs change with the doping level, modulating the intricate FS nesting conditions that are likely important for the spin-density wave ground state, which directly competes with superconductivity~\cite{Yi2014}. Hence the effect of the electron correlation with superconductivity is less direct. It has been proposed that orbital selective electron correlation may be a tuning parameter for superconductivity and unifies the iron-based superconductor phase diagrams, in which the spin-density wave order is accidental due to the details of the FS~\cite{DeMedici2014}. Here the metallic phases of \RFS, with only electron pockets at the BZ corner and no competing magnetic order, present a more direct pathway that showcases electron correlation and specifically bandwidth renormalization to be an important tuning parameter for superconductivity. More recently, a theoretical work using orbital-dependent exchange couplings has shown that in the case where \dxy~is selectively localized, there exists a pairing state where a node-less gap and the existence of a spin resonance indicative of sign-changing pairing symmetry may be reconciled~\cite{Nica2015}. Our observation of the strong orbital-selective renormalization in the RFSe superconductor confirms the applicability of this theory to these materials.

\begin{acknowledgments}
ARPES experiments were performed at the Advanced Light Source and the Stanford Synchrotron Radiation Lightsource, which are operated by the Office of Basic Energy Science, U.S. Department of Energy. The work at Berkeley and Lawrence Berkeley National Laboratory are supported by the Office of Science, Office of Basic Energy Sciences, U.S. Department of Energy, under Contract No. DE-AC02-05CH11231. The work at Berkeley is also supported by the Office of Basic Energy Sciences U.S. DOE Grant No. DE-AC03-76SF008. We also acknowledge support from NBRPC-2012CB821400 and NSFC11275279. The work at Stanford is supported by the DOE Office of Basic Energy Sciences, Division of Materials Science.
\end{acknowledgments}

%

\end{document}